\documentclass[aps,prl,twocolumn,superscriptaddress,showpacs]{revtex4-1}

\usepackage{amsmath}
\usepackage{graphicx}
\usepackage{bm}
\usepackage[T1]{fontenc}
\usepackage[latin1]{inputenc}
\usepackage{times}
\usepackage{amssymb}

\begin{document}

\title{Spatiotemporal coherent control of light through a multiply scattering medium with the Multi-Spectral Transmission Matrix}

\author{Mickael Mounaix} \email{mickael.mounaix@lkb.ens.fr} \affiliation{Laboratoire Kastler Brossel, ENS-PSL Research University, CNRS, UPMC Paris Sorbonne, Coll\`{e}ge de France, 24 rue Lhomond, 75005 Paris, France } 
\author{Daria Andreoli}  \affiliation{Laboratoire Kastler Brossel, ENS-PSL Research University, CNRS, UPMC Paris Sorbonne, Coll\`{e}ge de France, 24 rue Lhomond, 75005 Paris, France } \affiliation{ESPCI ParisTech, PSL Research University, CNRS, Institut Langevin, 1 rue Jussieu, F-75005 Paris, France}
\author{Hugo Defienne} \affiliation{Laboratoire Kastler Brossel, ENS-PSL Research University, CNRS, UPMC Paris Sorbonne, Coll\`{e}ge de France, 24 rue Lhomond, 75005 Paris, France }
\author{Giorgio Volpe}  \affiliation{Laboratoire Kastler Brossel, ENS-PSL Research University, CNRS, UPMC Paris Sorbonne, Coll\`{e}ge de France, 24 rue Lhomond, 75005 Paris, France } \affiliation{Department of Chemistry, University College London, 20 Gordon Street, London WC1H 0AJ, UK }
\author{Ori Katz}  \affiliation{Laboratoire Kastler Brossel, ENS-PSL Research University, CNRS, UPMC Paris Sorbonne, Coll\`{e}ge de France, 24 rue Lhomond, 75005 Paris, France } \affiliation{ESPCI ParisTech, PSL Research University, CNRS, Institut Langevin, 1 rue Jussieu, F-75005 Paris, France}
\author{Samuel Gr\'{e}sillon} \affiliation{ESPCI ParisTech, PSL Research University, CNRS, Institut Langevin, 1 rue Jussieu, F-75005 Paris, France}
\author{Sylvain Gigan}  \affiliation{Laboratoire Kastler Brossel, ENS-PSL Research University, CNRS, UPMC Paris Sorbonne, Coll\`{e}ge de France, 24 rue Lhomond, 75005 Paris, France } 

\begin{abstract}

We report broadband  characterization of the propagation of light through a multiply scattering medium by means of its Multi-Spectral Transmission Matrix. Using a single spatial light modulator, our approach enables the full control of both spatial and spectral properties of an ultrashort pulse transmitted through the medium. We demonstrate spatiotemporal focusing of the pulse at any arbitrary position and time with any desired spectral shape. Our approach opens new perspectives for fundamental studies of light-matter interaction in disordered media, and has potential applications in sensing, coherent control and imaging.
\end{abstract}

\pacs{}

\maketitle

Propagation of coherent light through a scattering medium produces a speckle pattern at the output~\cite{goodman1976some}, due to light scrambling by multiple scattering events~\cite{sebbah2012waves}. 
The phase and amplitude information of the light are spatially mixed, thus limiting resolution, depth and contrast of most optical imaging techniques. Ultrashort pulses, generated by broadband mode-locked lasers, are very useful for multiphotonic imaging and non-linear physics~\cite{oheim2001two,denk1990two,brabec2000intense}. In the temporal domain, an ultrashort pulse is temporally broadened during propagation in a scattering medium, due to the long dwell time within it~\cite{bruce1995investigation,johnson2003time}, which therefore limits its range of applications. 
 
However, this scattering process is linear and deterministic. Therefore, one can control the input wavefront to design the output field.
 In this respect, spatial light modulators (SLMs) offer more than a million degrees of freedom to control the propagation of coherent light. These systems have played an important role in the development of wavefront shaping techniques to manipulate light in complex media. Iterative optimization algorithm~\cite{vellekoop2007focusing,vellekoop2008universal,vellekoop2008phase} and phase conjugation methods~\cite{papadopoulos2012focusing,yaqoob2008optical} have been proposed to focus light at a given output position, an essential ingredient for imaging. An alternative method for  light control is the optical transmission matrix (TM). The TM is a linear operator that links the input field (SLM) to the output field (CCD camera)~\cite{beenakker1997random,popoff2010measuring}. The measurement of the TM allows imaging through~\cite{popoff2010image} or inside a scattering medium~\cite{chaigne2014controlling}, and potentially access mesoscopic properties of the system~\cite{PhysRevLett.111.063901}.

The possibility of shaping the pulse in time is also essential for coherent control~\cite{cruz2004use}. Temporally, photons exit  a scattering medium at different times, giving rise to a broadened pulse at its output~\cite{tomita1995observation, johnson2003time}. Temporal spreading of the original pulse is characterized by a confinement time $\tau_m$~\cite{curry_direct_2011} related to the Thouless time~\cite{thouless1974electrons}. Equivalently, from a spectral point of view, the scattering medium responds differently for distinct spectral components of an ultrashort pulse, with a spectral correlation bandwidth $\Delta \omega_m \propto 1/\tau_m$, giving rise to a very complex spatio-temporal speckle pattern ~\cite{andreoli2015MSTM,mosk2012controlling,small2012spectral, van2011frequency}. With a single SLM, one can manipulate spatial degrees of freedom to adjust the delay between different optical paths. Therefore spatial and temporal distortions can be both compensated using wavefront shaping techniques. This approach allows the temporal compression of an ultrashort pulse at a given position by iteratively optimizing the input wavefront~\cite{aulbach_control_2011,paudel2013focusing,katz2011focusing}, or alternatively by using digital phase conjugation~\cite{morales2015delivery}. Another technique consists in shaping only the spectral profile of the pulse at the input, to compensate the temporal distortion induced by the medium~\cite{mccabe_spatio-temporal_2011}. Equivalently to a spectral approach, the measurement of a time-resolved reflection matrix~\cite{PhysRevLett.111.243901, kang2015imaging} enables light delivery at a given depth of the scattering medium.

Recently, the measurement of a TM of the medium for several wavelengths, the Multi-Spectral Transmission Matrix (MSTM)~\cite{andreoli2015MSTM}, allowed for both spatial and spectral control at any position in space using a single SLM. In~\cite{andreoli2015MSTM}, since the spectral phase relation between different matrices was unknown, deterministic temporal control was still elusive. Here, we introduce the MSTM formalism including the spectral phase relation between the different frequency responses of the medium. This additional information gives access to a full spatio-temporal control of an ultrashort pulse propagating in the disordered medium. We demonstrate deterministic spatiotemporal focusing and enhanced excitation of a non-linear process. Beyond this dispersion compensation process, we also deterministically shape the temporal profile of the output pulse.

\vskip 0.25cm

\begin{figure*}
\includegraphics[width=1\textwidth]{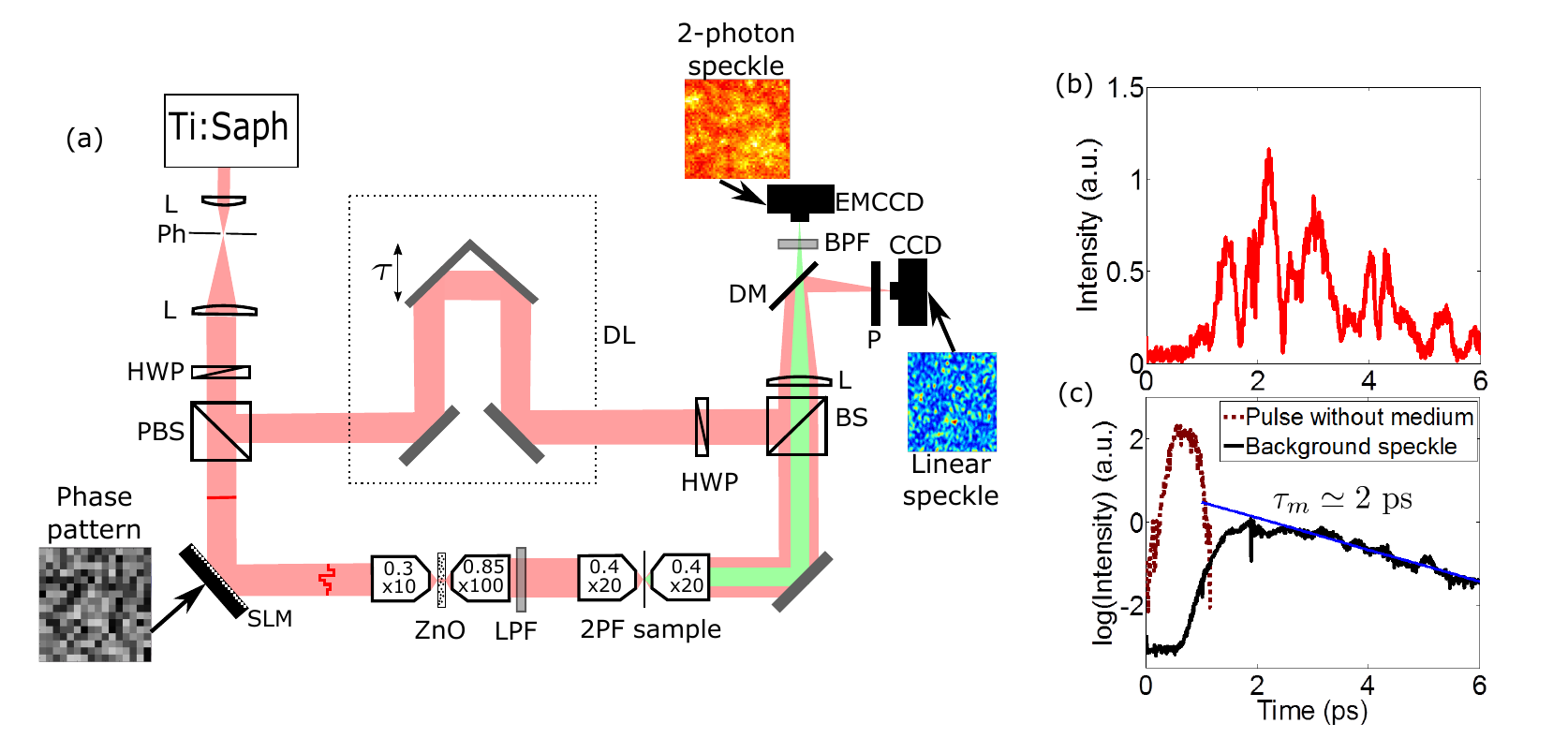}
\caption{ \label{manip}(Color online)  Principle of the experiment. (a) Experimental setup. The laser (ultrashort pulse 110 fs at FWHM) is expanded and illuminates the SLM which is conjugated with the back focal plane of a microscope objective. The scattering medium is a thick (around 100 $\mu m$) sample of ZnO nanoparticles, placed between two microscope objectives. The output plane is imaged on the two-photon fluorescence (2PF) sample. The long-pass filter (LPF) allows one to eliminate the autofluorescence of the ZnO. A dichroic mirror (DM) separates the two-photon fluorescence signal (imaged with an EMCCD) from the linear signal (imaged with a CCD). Lens (L), pinhole (Ph), half plate wave (HPW), polarized beam-splitter (PBS), delay line (DL), beam splitter (BS), polarizer (P), band pass filter (BPF).
(b) Temporal speckle from a single spatial speckle grain, obtained with an Interferometric Cross-Correlation (ICC) measurement.
(c) ICC of the original pulse in the sample arm without the medium (dotted line),  characterized by a time-width of 520 fs at FWHM.  ICC of the transmitted pulse through the medium, averaged over 100 different spatial positions of the  speckle  (solid line), characterized by a decay time of  $\simeq2$~ps (blue line). }
\end{figure*}

First, we need to describe the propagation of an ultrashort pulse through the scattering medium, i.e. measure the MSTM. The ratio between the spectral bandwidth of the ultrashort pulse $\Delta \omega_L$ and $\Delta \omega_m$ gives the number of independent spectral degrees of freedom $N_{\omega}$. It corresponds to the number of  monochromatic TM that one needs to measure to completely describe  both spatially and spectrally the propagation of the broadband signal. In~\cite{andreoli2015MSTM}, the reference signal in the measurement of the MSTM was a co-propagative speckle, as in~\cite{popoff2010TM}. Since this reference speckle is also $\lambda$-dependent, the spectral phase remained undetermined. Nonetheless, the manipulation of the input field with this incomplete MSTM allows using the scattering medium as a very complex optical component, such as a lens or a grating~\cite{andreoli2015MSTM}. Achieving full control of the pulse in the time domain requires the knowledge of the relative phase relation between the wavelengths of the pulse at the output of the medium. Here, the reference field is a plane wave of known phase introduced by an external reference arm. This reference field is common to all the individual monochromatic TMs: the relative phase between the different wavelengths of the output pulse is then accessible in every spatial position at the output. The output field reads:

\begin{equation} \label{MSTMeq}
E_j^{out}=\sum_{i=1}^{N_{\text{SLM}}} \sum_{k=1}^{N_{\omega}}{h_{jik}e^{i\varphi_{jk}}E_i^{in}(\lambda_k})
\end{equation}
where $E_j^{out}$ represents the value of the output field at the j-th pixel of the CCD camera, $E_i^{in}(\lambda_k)$ the value of the input field at the i-th SLM pixel at wavelength $\lambda_k$, $h_{jik}e^{i\varphi_{jk}}$ the coefficients of the MSTM with $\varphi_{jk}$ the spectral phase component.

Fig.~\ref{manip} illustrates the experimental setup. A Ti:Saph laser source (MaiTai, Spectra Physics) produces a 110fs ultrashort pulse, centered at 800nm with a spectral bandwidth of 10 nm FWHM. The same laser can also be used as a tunable monochromatic laser in the same spectral range. The phase-only SLM (LCOS-SLM, Hamamatsu X10468) is used in reflexion, subdivided in 32$\times$32 macropixels, which is a good compromise between duration of the measurement of the MSTM and efficiency of the focusing process. The scattering medium is a thick layer of ZnO nanoparticles randomly distributed on a glass slide, placed between two microscope objectives. Most experiments are performed on an approximately 100 $\mu$m thick sample,  characterized by its spectral correlation bandwidth $\Delta \lambda_m \simeq$ 0.5 nm (measurement done in the same way than~\cite{andreoli2015MSTM}), equivalent to $\Delta \omega_m \simeq$ 3 THz.  The two paths have the same optical length: when the laser is generating an ultrashort pulse, both the ultrashort reference pulse and the stretched one are overlapped in space and time. The output plane of the scattering medium is imaged on a two-photon fluorescent sample, which is a powder of fluorescein diluted in ethanol, inside a glass capillary (CM Scientific, 20 $\mu$m $\times$ 200 $\mu$m $\times$ 5 cm). The excitation of this non-linear process will be used to differentiate  the spatiotemporal  from the spatial focusing. A longpass filter is placed between the two sets of microscope objectives to eliminate some residual autofluorescence of the scattering medium. A CCD camera records the linear output intensity, and is used to measure the MSTM. The 2-photon signal is recorded with an EMCCD camera, placed after a dichroic mirror to separate the linear signal from the fluorescence. 
In a first step, the laser is set to CW, and we measured the MSTM  for $N_{\omega} = $21 wavelengths in a 13-nm spectral window centered at 800 nm. In essence, for each wavelength, we display a series of phase patterns on the SLM, and for each input pattern, phase stepping holography with the external reference arm wave permits to recover the complex output field. The MSTM is deduced from this set of measurement. Once the MSTM has been measured, the laser is set back to pulsed operation.
The CCD camera integrates the output speckle over time and only recovers its spatial fluctuations. The temporal shape of the speckle is retrieved with an Interferometric Cross-Correlation (ICC) measurement between the stretched pulse and an ultrashort reference by scanning the reference arm. The temporal envelope of this signal is then retrieved by applying a low pass filter in the Fourier domain~\cite{monmayrant2010newcomer}.

\begin{figure}[htbp]
\includegraphics[scale=1]{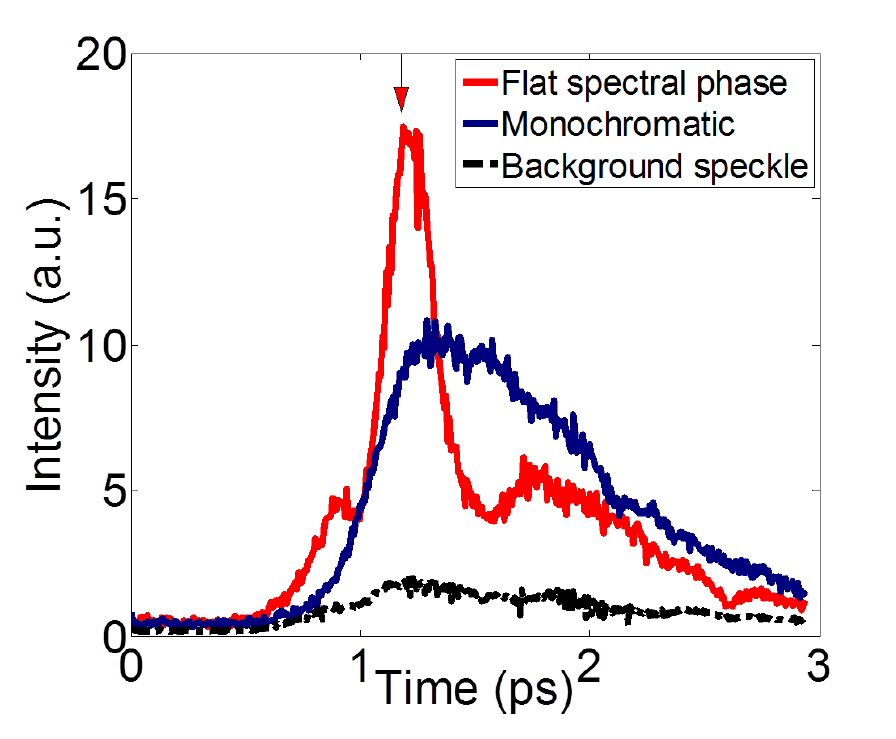}
\caption{\label{STfoc} (Color online) Spatio-temporal focusing with MSTM, averaged over 9 different spatial positions at the output. Spatio-temporal focusing is achieved by imposing a flat spectral phase at the output, whereas spatial focusing is obtained by focusing only the central wavelength of the pulse. The MSTM was measured at a delay-line position corresponding to the top arrow. }
\end{figure}

The temporal envelope of an output signal recorded at one given spatial position  (one speckle grain)  is shown on Fig.~\ref{manip}b. The measured output pulse is strongly elongated compared to the initial signal and it shows a very complex temporal structure. Averaging over 100 different output positions allows to recover a smooth shape (Fig.~\ref{manip}c). Its linear decay rate gives access to the confinement time $\tau_m \simeq 2$ ps corresponding to the time of flight distribution of the photons inside the medium. The value obtained is in agreement with the spectral correlation bandwidth $\Delta \omega_m$ independently measured. This averaged envelope can be compared to the envelope of the pulse reconstructed without the scattering medium. The ultrashort pulse in the reference arm is transform limited at 110 fs, whereas the same pulse after propagation in the sample arm is dispersed to a time-width of 520 fs at FWHM (mainly because of the microscope objectives).

At this stage, we want to use the MSTM to generate a particular temporal profile at a given output spatial position. The input pattern to focus a given wavelength $\lambda_k$ at a given position can be simply obtained by phase-conjugating the corresponding line of the monochromatic TM($\lambda_k$)~\cite{popoff2010TM}, thus generating a focus without any temporal compression. To extend this idea to the temporal domain, we can generalize this principle.  It is possible, using a single SLM pattern, to focus multiple wavelengths $\lambda_k$  simultaneously at a given position, with a well-defined spectral phase $\theta_k$. The corresponding solution can be obtained by algebraically summing all the individual patterns. Since the SLM is phase-only, the optimal phase-pattern to display is simply the argument of the solution. For example, to achieve spatiotemporal focusing, each frequency is focused into a specific output spatial position, while simultaneously ensuring that their relative phases are equal: $\theta_1=\dots=\theta_{N_\omega}$.

\begin{figure}[htbp]
\includegraphics[scale=1]{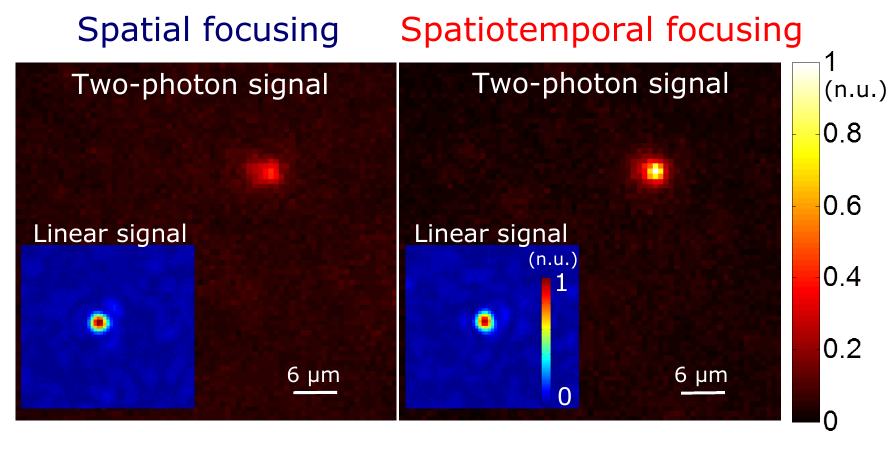}
\caption{\label{NL} (Color online) Comparison of 2-photon excitation for spatial (monochromatic phase conjugation) and spatiotemporal focusing, obtained with a thinner scattering sample of ZnO ($\tau_m\simeq$ 500 fs). Inset : corresponding CCD signal. Integration time identical for both excitations : 2s for the 2-photon and 1ms for the linear. Scale bars normalized to the maximum intensity of spatiotemporal focusing. We observe a two-fold enhancement of the non-linear signal for the same average intensity when performing spatiotemporal focusing.}
\end{figure}

The main result is presented in Fig.~\ref{STfoc}, where we demonstrate temporal compression of the pulse. The presented temporal profiles are retrieved with an ICC measurement.  All data are averaged over 9 different spatial positions at the output for better visibility. Imposing a flat spectral phase (i.e. the same spectral phase for all wavelengths) at a chosen spatial position at the output, we observe spatiotemporal focusing at a time determined by the delay line position where the MSTM was previously measured, corresponding to the top arrow  in Fig.~\ref{STfoc}. Temporal compression of the output pulse is obtained almost to its Fourier limited time-width (120 fs at FHWM). The shape of the pulse around the peak is not due to artifacts but is due to the square spectral window used in the MSTM measurement, giving a cardinal sine form with an expected rebound at 400fs from the arrival time of the pulse.  As a comparison, spatial focusing is also demonstrated by focusing only the central wavelength of the output pulse~\cite{supp_mat3}. The observed intensity is higher than the averaged background speckle because the pulse is spatially focused, but temporal compression is missing.

To unequivocally demonstrate the temporal compression, we complement our linear ICC characterization by a  two-photon fluorescence measurement. The total intensity of the two-photon fluorescent signal is proportional to the square of the excitation intensity, and also to the inverse of its time-width \cite{zipfel2003nonlinear}. Therefore, with an equivalent spatial focusing, a temporal compression corresponds to a higher two-photon signal. 
In Fig.~\ref{NL}, we show the two fluorescent signals related to input wavefronts that give either spatial-only (signal-to-background ratio (SBR): 8.5) or spatiotemporal focusing (SBR: 19.6) at the output of the medium, with a thinner scattering sample of ZnO (characterized by a measured  $\Delta \lambda_m\simeq$ 2 nm, requiring the use of 6 monochromatic TM) to increase the two-photon signal contrast. The signal over noise ratio of the two-photon signal is about 2.5 times higher when the light is spatiotemporally focused compared to the case where the light is spatially focused, with a similar linear signal and same focus spot size in both cases~\cite{supp_mat1}.

\begin{figure}[htbp]
\includegraphics[scale=1]{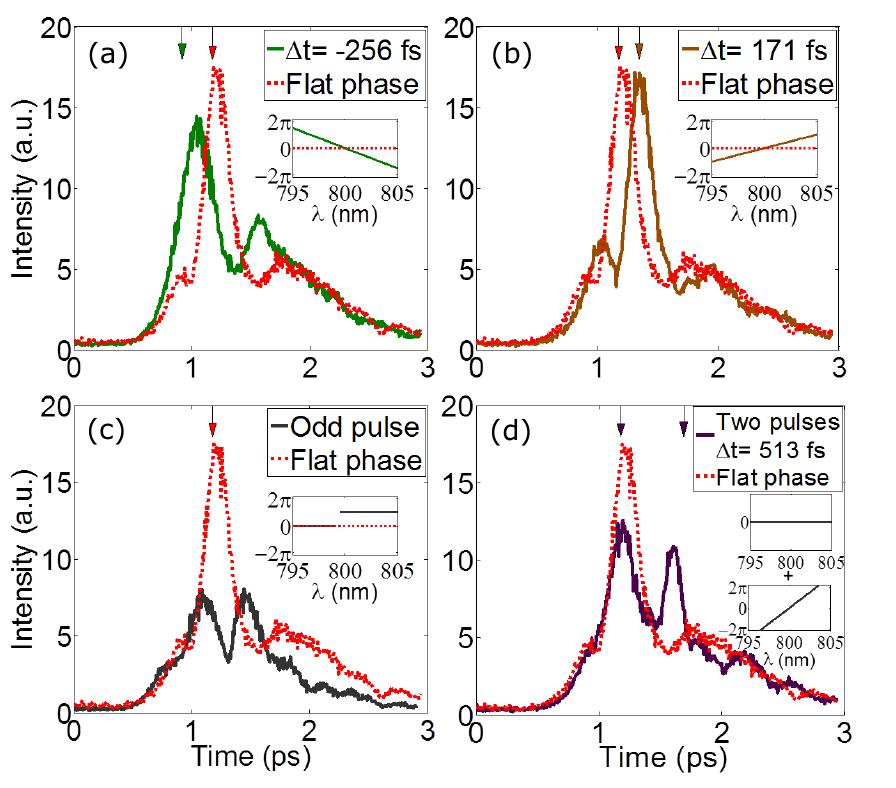}
\caption{\label{ST2} (Color online) Any suitable spectral shape can be obtained. (a) A negative spectral phase ramp advances the arrival time of the pulse. (b) A positive spectral phase ramp delays the pulse. (c) A $\pi$-phase step in the spectral domain creates an odd pulse. (d) The sum of a flat phase pulse and a spectral phase ramp pulse  gives two pulses with a controlled delay at the same output spatial position. Inset : spectral phase profile applied to the signals. Top arrows indicate the expected arrival time of the pulse.}
\end{figure}

The MSTM gives also access to a more sophisticated spectral shape. The control of this information allows any kind of spectral shape at the output of the scattering medium~\cite{weiner2000femtosecond}, without any additional measurement. In essence, the scattering medium in conjunction with the spatial light modulator can be used as a pulse shaper. Results of temporal control on the output signal are shown in Fig.~\ref{ST2}~\cite{supp_mat2}. In  Fig.~\ref{ST2}a and  Fig.~\ref{ST2}b a linear spectral phase relation $\theta$ between the wavelengths of the pulse is imposed, thus advancing or delaying the arrival time of the pulse. By tuning the slope of the imposed spectral phase ramp~\cite{weiner2000femtosecond}, the ultrashort pulse can be temporally shifted with a controllable delay: 

\begin{equation}
\tau = -\frac{\delta \phi \text{ } \lambda_0^2}{2\pi  c \text{ } \delta \lambda}
\end{equation}
where $\delta \phi$ is a phase difference  between the first and the last wavelength in a spectral interval $\delta \lambda$ centered around $\lambda_0$, and $c$ is the speed of light. The predicted arrival time of the pulse is indicated by the top arrows on each plot. The ultrashort pulse is focused in the same output spatial position, but its arrival time is changed. Two other important examples are demonstrated. Firstly, a $\pi$-phase step between the two halves of the spectrum induces a pulse with a dip, also called an odd pulse~\cite{weiner1987picosecond,weiner1992programmable}, which can be extremely useful for coherent control~\cite{meshulach1999coherent}. The corresponding temporal profile is presented in  Fig.~\ref{ST2}c.
Secondly, a linear superposition of a flat phase pulse and a spectral phase ramp pulse on the SLM at the same spatial position allows the arrival of two pulses with a controllable delay as the two solutions are incoherent with themselves because of the temporal speckle. This could  allow pump-probe excitation. As an example,  we show  in Fig.~\ref{ST2}d two pulses separated by $\Delta$t = 513 fs.

In all these experiments, the efficiency of the focusing, i.e. the signal to background ratio of the obtained spatiotemporal shape relative to the background speckle, scales linearly with the number of controllable degrees of freedom, here the number of controlled segments on the SLM~\cite{aulbach_control_2011}. It also depends on the number of different spatial targets~\cite{vellekoop2007focusing, popoff2010TM}, temporal targets and on the number of spectral degrees of freedom $N_{\omega}$~\cite{lemoult2009manipulating}. 

In conclusion, our setup allows us to access both the spatial and spectral phase information of the  MSTM, thus using the multiply scattering medium in conjunction with the SLM as a lens and a pulse shaper.  It is also conceivable to focus two pulses at two different times at two different spatial positions, or to control the duration of the pulse by adding a quadratic phase relation between the frequencies. As these examples indicates, any temporal shape is achievable, with a resolution given by the temporal duration of the pulse, over a temporal interval related to the confinement time of the medium and limited spatially by the speckle grain size. It can be  achieved using only spatial degrees of freedom of a single SLM. This temporal control  could enable  coherent quantum control~\cite{meshulach1998coherent}, or the excitation of localized nano-objects, and open interesting perspectives for the study of light-matter interaction, non-linear imaging in multiply scattering media, and more fundamental insights  such as light transport properties~\cite{wang2011transport}.

\begin{acknowledgments}
The authors would like to thank Thomas Chaigne for inspiring discussions. This work was funded by the European Research Council (grant no. 278025)
\end{acknowledgments}

\pagebreak
\widetext
\begin{center}
\textbf{\large Supplemental Materials: Spatiotemporal coherent control of light through a multiply scattering medium with the Multi-Spectral Transmission Matrix}
\end{center}
\setcounter{equation}{0}
\setcounter{figure}{0}
\setcounter{table}{0}
\setcounter{page}{1}
\makeatletter
\renewcommand{\theequation}{S\arabic{equation}}
\renewcommand{\thefigure}{S\arabic{figure}}
\renewcommand{\bibnumfmt}[1]{[S#1]}
\renewcommand{\citenumfont}[1]{S#1}

\section{Part 1: Algorithm for spectral shaping using the Multi spectral Transmission Matrix}

In this section we present the algorithm we use to arbitrary shape the output pulse with the Multi Spectral Transmission Matrix (MSTM).

\subsection{A: Spatial only focusing (monochromatic focusing)}
 Firstly, let us introduce how to perform a spatial only focusing process of the output pulse at a given position. For this purpose, we focus only one given specific wavelength $\lambda_k$ of the output pulse into one specific output position. We consider the transmission matrix equation~\ref{eqTM}~\cite{popoff2010measuring} that reads:

\begin{equation} \label{eqTM}
E_{output}=H E_{in} 
\end{equation}
 with H= TM($\lambda_k$) the transmission matrix measured at wavelength $\lambda_k$. The input field  required to focus at a given spatial target output position is calculated using the transpose conjugate of H.

\begin{equation} \label{spatialfoc}
E_{in}= H^\dagger E_{target}
\end{equation}

where $E_{target}$ is a vector containing zeros everywhere except for a coefficient 1 positioned in the row associated to the targeted pixel on the CCD camera. \\

\begin{equation} \label{target_spatial}
E_{target} =  \left[ \begin{array}{c} 0 \\ \vdots  \\ 1 \\ \vdots  \\ 0 \end{array} \right] 
\end{equation}
The full algorithm is detailed below and illustrated in Fig.~\ref{spatial_foc_scheme}: 

\begin{figure}[htbp]
\begin{center}
\includegraphics[width=0.8\textwidth]{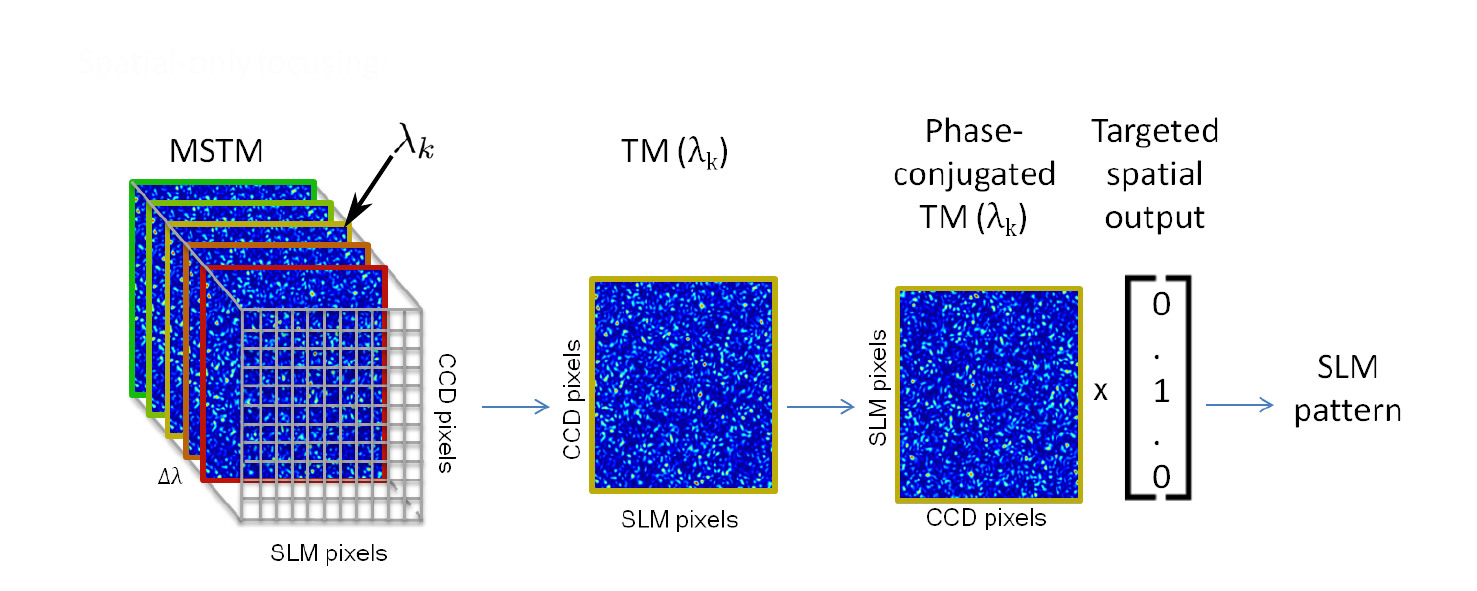}
\caption[0.5\textwidth]{Scheme of the algorithm used in the spatial only focusing process of an ultrashort pulse using the MSTM. Like in~\cite{popoff2010measuring}, the phase conjugation of one monochromatic transmission matrix enables to focus the corresponding wavelength of the output pulse.}
\label{spatial_foc_scheme}
\end{center}
\end{figure}

\begin{itemize}
\item (A1) We extract from the MSTM the monochromatic TM($\lambda_k$) corresponding to the wavelength $\lambda_k$~\cite{andreoli2015deterministic}.
\item (A2) We calculate the complex conjugate of the monochromatic matrix associated to this wavelength $\lambda_k$, and multiply it by the targeted spatial output $ E_{target}$.
\item (A3)  As the SLM is only able to modulate the phase of the input field, we display on the SLM the phase of the previously calculated solution.  
\end{itemize}

This algorithm is described with more details in~\cite{popoff2010measuring}. In particular, as mentionned in ref [23] of~\cite{popoff2010measuring}, using only the phase of the solution gives the optimal solution for concentrating energy in the focus. The resulting focus remains temporally broadened as only one wavelength of the output pulse is controlled.

\subsection{B: Arbitrary pulse shaping}

\begin{figure}[htbp]
\begin{center}
\includegraphics[width=0.8\textwidth]{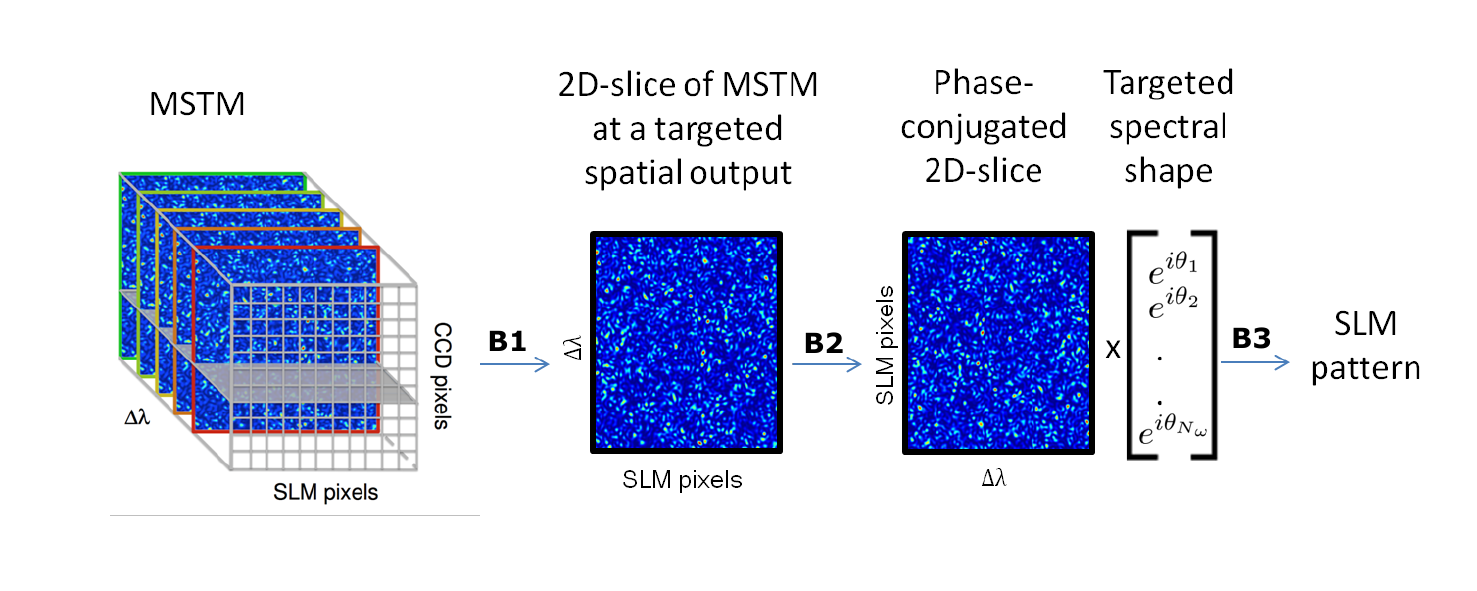}
\caption[0.5\textwidth]{Scheme of the algorithm used to arbitrary shape the output pulse at a given output position. The phase relation ($\theta_i$) between the $N_{\omega}$ wavelengths is imposed at the output of the scattering medium by the user, and the corresponding input is calculated. }
\label{arbitrary_shaping}
\end{center}
\end{figure}

We now introduce the algorithm used to spatio-temporally control the propagation of the pulse. For this purpose, we want the output pulse to have a specific spectral phase relation ($\theta_i$) between its different wavelengths at one given output position, the CCD pixel where we want to shape the pulse. Fig.~\ref{arbitrary_shaping} shows the algorithm used to compute the corresponding SLM pattern.

\begin{itemize}
\item (B1) We extract from the MSTM a 2D-\textsl{slice} at the corresponding CCD pixel (gray plan on first step of  Fig.~\ref{arbitrary_shaping}). This matrix-slice is denoted H and it contains the relation between the input field and the different wavelengths of the output pulse at this specific spatial output position. 
\item (B2) By analogy to Eq.~\ref{spatialfoc}, we use the transpose conjugate of H to determine the spatial shape of the input field. In this case, the targeted output field $E_{target}$ is a vector $E_{target}(\lambda)$ giving the desired phase relation ($\theta_i$)  between the different wavelengths of the output pulse:

\begin{equation} \label{spectral_phase}
E_{target}(\lambda)= \left[ \begin{array}{c} e^{i \theta_{1}} \\ e^{i \theta_{2}} \\ \vdots \\ e^{i \theta_{N_{\omega}}} \end{array} \right] 
\end{equation}

\item (B3)  As the SLM is only able to modulate the phase of the input field, we display on the SLM the phase of the previously calculated solution.  
\end{itemize}

\subsection{C: Examples of pulse shaping}

In this section we give some details about the pulse shaping results presented in Fig. 4 of the manuscript. Arbitrary spectral shaping is achievable since we can set at will the spectral phase ($\theta_i$) at any specific position at the output.

\subsubsection{Spatio-temporal focusing}
The simplest way to achieve spatio-temporal focusing at the output is to impose a flat phase profile.  Therefore, the phase relation ($\theta_i$) of Eq.~\ref{spectral_phase}  reads $\theta_1=\dots=\theta_{N_{\omega}}$.

\subsubsection{Delaying or advancing the pulse}
A linear spectral phase relation ($\theta_i$) enables spatio-temporal focusing of the output pulse pulse at a specific delay time relative to the flat spectral phase. . 
The slope of ($\theta_i$) imposes the arrival time of the output pulse:

\begin{equation}
\theta_i = \dfrac{(i-1)-\dfrac{N_{\omega}}{2}}{N_{\omega}} \delta \phi
\end{equation}
with $i$ the index of each individual $N_{\omega}$ wavelength over a spectral interval $\delta \lambda$ centered around $\lambda_0$, varying from 1 to $N_{\omega}$, and $\delta \phi$ the phase difference between the first and the last wavelength. The delay $\tau$ in comparison with an imposed flat phase reads:

\begin{equation}
\tau = -\frac{\delta \phi \text{ } \lambda_0^2}{2\pi  c \text{ } \delta \lambda}
\end{equation}
with $c$ the speed of light. Therefore, by tuning $\delta \phi$ and its sign, the arrival time of the output pulse is controllable.

In the experiment presented in Fig.4a and Fig.4b of the manuscript, we chose the phase ramps to impose $\tau_a= -256$ fs and $\tau_b= 171$ fs. 

\subsubsection{Double pulses with controllable delay}

In the previous section, we demonstrated how to focus the output pulse at a specific position in space and time. We now exploit the linearity of the system to focus the output pulse at one given spatial position but at two different times. For this purpose, the targeted output field reads:

\begin{equation} 
E_{target}(\lambda)= \underbrace{\left[ \begin{array}{c} e^{i \theta_{1}} \\ \ldots \\ e^{i \theta_{N_{\omega}}} \end{array} \right]}_{\tau_1} + \underbrace{\left[ \begin{array}{c} e^{i \theta'_{1}} \\ \ldots \\ e^{i \theta'_{N_{\omega}}} \end{array} \right]}_{\tau_2}
\end{equation} 
where the linear phase relation between $[\theta_{1} \ldots \theta_{N_{\omega}}]$ corresponds to a focus at time $\tau_1$ and $[\theta'_{1} \ldots \theta'_{N_{\omega}}]$ at time $\tau_2$.

The corresponding SLM pattern is determined by the algorithm presented in Fig.~\ref{arbitrary_shaping}. The intensity of each pulse is lowered by a factor 2 compared to the situation when focusing individually each pulse, since the number of degrees of freedom is the same.

In the experiment presented in Fig.~4d of the manuscript, we chose the phase ramps to impose $\tau_1= 0$ fs (flat phase) and $\tau_2= 513$ fs. 

\subsection{D: Comparison between monochromatic and polychromatic control for spatial-only focusing}

In the manuscript, we exploit the full MSTM  with a common reference field to  control the spectral phase $(\theta_i)$, and thus the temporal profile of the output pulse. 

In order to compare with situations where the pulse is not temporally recompressed, Fig.~2 and Fig.~3 show comparisons with monochromatic phase-conjugation, where only a narrow spectral band is focused, corresponding to the central wavelength, together with a spectral band around it given by the spectral correlation bandwidth of the medium. 

Here, in order to give a more complete picture, we compare this monochromatic spatial-only focusing with two alternative ways to generate a spatial-only focus using the full spectrum of the pulse:

\begin{itemize}
\item If the MSTM is measured as in~\cite{andreoli2015deterministic} (\textsl{i.e.} with a co-propagative speckle as a reference field), the relative phase between the different wavelengths is unknown,  as the reference field is $\lambda$-dependent. In this case, a polychromatic focusing as described in part B of this Supplementary Material corresponds to focusing all the wavelengths at a given spatial output position without controlling the spectral phase relation. It results in a spatial focus, but the temporal signal remains broadened. 
\item If the MSTM is measured with an external reference field, as in this paper, an equivalent result can be achieved by deliberately imposing a random spectral phase relation $(\theta_i)$ between the different spectral components being focused.  It also results in a spatial focus, but the temporal signal remains broadened. 
\end{itemize}

\begin{figure}[htbp]
\begin{center}

\includegraphics[width=0.53\textwidth]{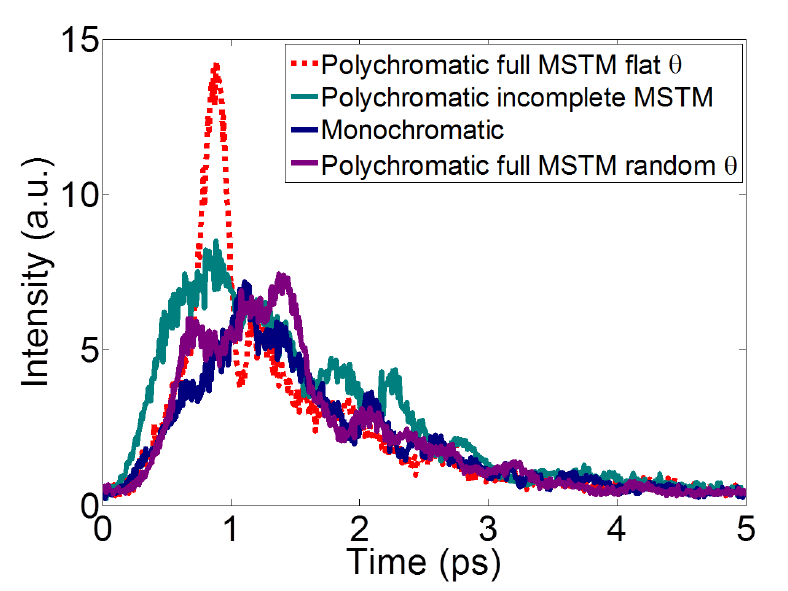}
\caption[0.5\textwidth]{Measurement of the temporal broadening of the foci obtained by different methods of spatial-only focusing: (magenta) spatial-only polychromatic focusing obtained with the full MSTM, setting  a random spectral phase relation $(\theta_i)$, (green) spatial-only focusing using the incomplete  MSTM measured with a co-propagating reference speckle, (blue) spatial-only focusing obtained  by phase conjugating only the central wavelength of the pulse (as in Fig.~2), and (red) comparison with spatio-temporal focusing, obtained with a flat spectral phase relation $(\theta_i)$ as in Fig.~2. All the plots are spatially averaged over 9 different spatial positions at the output. All measurements are done \textsl{via} interferometric cross correlation as described in Fig. 2 of the manuscript.}
\label{polychromatic}
\end{center}
\end{figure}

We show in  Fig.~\ref{polychromatic} the temporal profiles of the foci, obtained with the three methods.  Fig.~\ref{polychromatic} shows that the three approaches  give  equivalent results in terms of temporal broadening. In all  cases, the average temporal profile of the focus corresponds to the natural confinement time of the medium.  Therefore,  spatial-only focusing can be equivalently achieved  with a polychromatic focusing (using the complete or incomplete MSTM) or with a monochromatic phase-conjugation. In the manuscript, for simplicity's sake, we chose to compare the temporally compressed output pulse to the  monochromatic case only. 

\section{Part 2: 2-photon enhancement with spatio-temporal focusing }

In the manuscript, we show the possibility to enhance a 2-photon fluorescence process as the output pulse is temporally compressed. In this section, we compare the enhancement of both linear (at $\lambda$ =800 nm) and the 2-photon signal (at $\lambda \simeq$ 400-600 nm) in the case of spatial but temporally broadened (spatial only, monochromatic in the manuscript) focusing and spatio-temporal focusing.

In spatial only and spatio-temporal focusing, the linear images measured with a CCD camera show an equivalent intensity peak and signal-to-background ratio. Indeed, 
\begin{itemize}
\item In the monochromatic spatial only focusing process, we use N degrees of freedom, the N independent SLM segments controlled, to focus a speckle pattern at a specific wavelength $\lambda_k$ within a set of $N_{\omega}$ wavelengths. For this specific wavelength  $\lambda_k$, the signal-to-background ratio is then approximately on the order of N. The other $N_{\omega}$-1 speckles are summing incoherently and they simply contribute to increase the level of the background by approximately $N_{\omega}$. At the end, the signal-to-background ratio observed on the linear image is on the order of N/$N_{\omega}$. 
\item 	In the spatio-temporal focusing  focusing process, we use the same N degrees of freedom to focus the output pulse both spatially and temporally. From a spectral point of view, this is equivalent to focus spatially each of the individual $N_{\omega}$ monochromatic speckles created at the $N_{\omega}$ different wavelengths. In average, each speckle is then spatially focused by using N/$N_{\omega}$ degrees of freedom. This gives a signal-to-background ratio on the order of N/$N_{\omega}$ in each speckle. Finally, the incoherent sum of these speckles generates a linear image with the same signal-to-background ratio N/$N_{\omega}$. The reasoning is identical for a polychromatic focusing, as it is independent of the imposed spectral phase relation $(\theta_i)$. 
\end{itemize}

\begin{figure}[htbp]
\begin{center}
\includegraphics[width=0.6\textwidth]{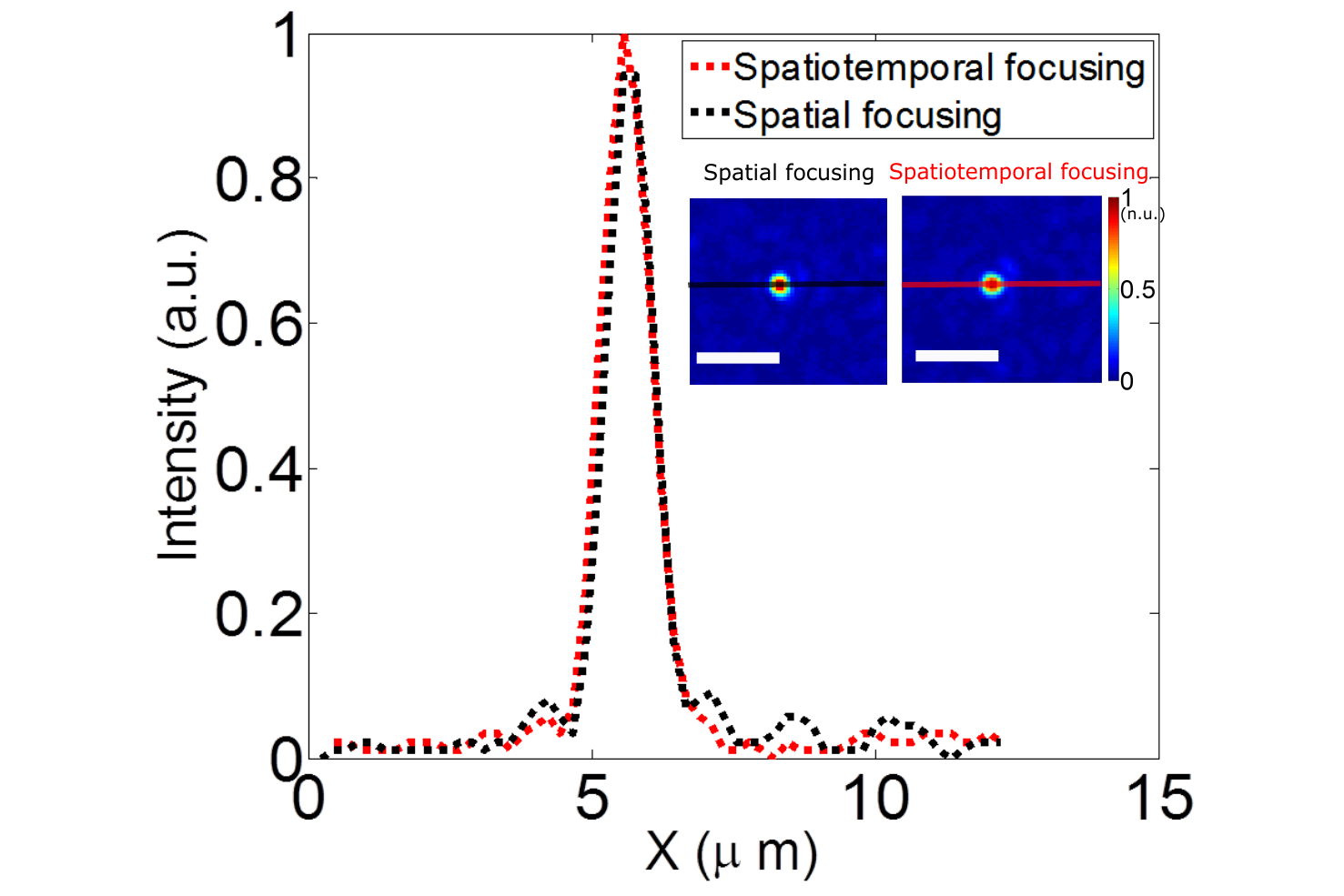}
\caption[0.5\textwidth]{Profiles  pictures of linear focus spots of spatial and spatio-temporal focusing, taken along the colored line of the inset. The two foci have the same size (approximately 1 $\mu m$), which is the size of a  speckle grain. Inset: CCD image of spatial (left) and spatio-temporal (right) focusing. Scale bar is 5 $\mu m$}
\label{linear}
\end{center}
\end{figure}

In the linear images presented on Fig.~3 of the manuscript, and on the inset of Fig.~\ref{linear} of this Supplementary Material, the signal-to-background ratios for both the spatial only focusing and the spatio-temporal focusing are respectively equal to 42 and 45. Therefore, for the same input power, the two linear images are nearly equivalent. \\

On the corresponding 2-photon fluorescence images, taken with the other camera, we measured signal-to-background ratios for both the spatial only focusing and the spatio-temporal focusing respectively of 8.5 and of 19.6. While the increase in SNR of the two-photon signal is an unequivocal sign of temporal recompression, one can see that the value of the SNR is still far from the theoretical value for a quadratic process. This can be attributed to the two-photon screen finite thickness (20 $\mu m$), which means that the camera records the two photon fluorescence corresponding to several speckle grain planes at the same time in our setup, therefore increasing the background and also enlarging the apparent two-photon focus size.\\

Besides, the size of the linear focusing spot at the output is diffraction-limited. In our case, the spot size in both the spatial only or the spatio-temporal focusing configurations are the same because of the narrow relative spectral bandwidth. Indeed,  $\delta \lambda / \lambda_0  = 0.02 \ll1 $ of the input pulse ($\lambda_0$ = 800nm is the central wavelength of the pulse and $\delta \lambda$ =10nm its spectral bandwidth). A similar approach with the 2-photon fluorescence signals leads to an identical 2-photon focusing spot size in the case of a spatial or a spatio-temporal focusing.

\begin{figure}[htbp]
\begin{center}
\includegraphics[width=0.6\textwidth]{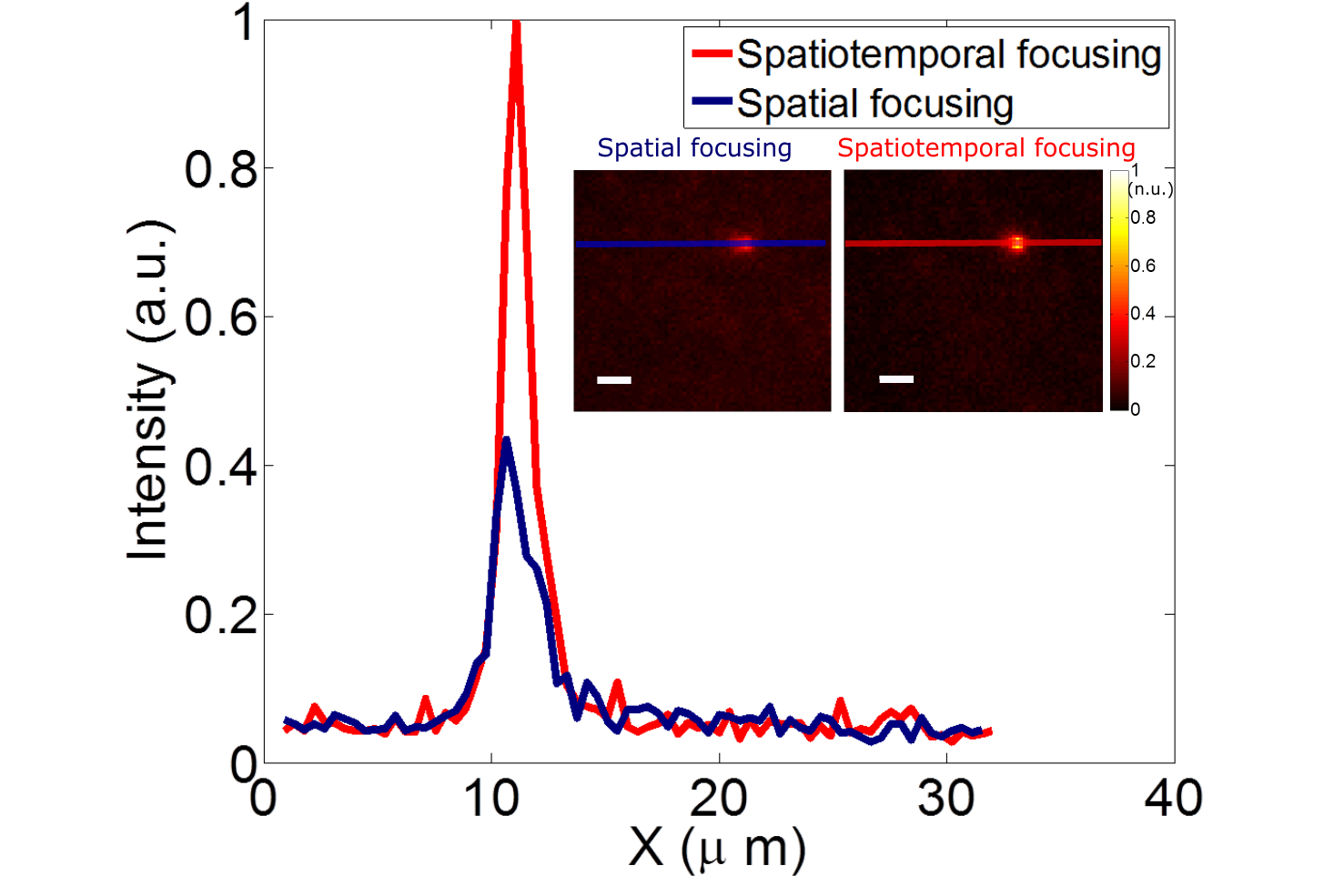}
\caption[0.5\textwidth]{Profiles  pictures of 2-photon foci spots of spatial and spatio-temporal focusing, taken along the colored line of the inset. The two foci have the same size (approximately 2 $\mu m$). Inset: 2-photon excitation under spatial (left) and spatio-temporal (right) focusing. Scale bar is 5 $\mu m$}
\label{2photon}
\end{center}
\end{figure}

In Fig.~\ref{linear} and  Fig.~\ref{2photon}, we have plotted the 1D-profile respectively of linear images and 2-photon images, in both configuration of spatial and spatio-temporal focusing. As expected, spatial and spatio-temporal focusing spots have the same linear spot size (approximately 1 $\mu m$ at FWHM), and also the same 2-photon spot size (approximately 2 $\mu m$ at FWHM).

\end{document}